\documentclass[reprint,amsmath,amssymb,aps,prr]{revtex4-1}
\usepackage{mathtools,cancel}
\usepackage{graphicx,amssymb}
\usepackage{dcolumn}
\usepackage{bm}

\usepackage{physics}
\usepackage{color}

\def\vr{\vec{r}}

\begin{document}

\title{A general model for vegetation patterns including rhizome growth}

\author{Daniel Ruiz-Reyn\'es$^1$, Francesca Sch\"onsberg$^{1,2}$, Emilio Hern\'andez-Garc\'ia$^1$ and Dami\`{a} Gomila$^1$}

\date{\today}%

\affiliation{$^1$IFISC (CSIC-UIB). Instituto de F\'isica Interdisciplinar y Sistemas Complejos,
E-07122 Palma de Mallorca, Spain \\
$^2$SISSA--Scuola Internazionale Superiore di Studi Avanzati, 34136 Trieste, Italy}

\begin{abstract}
Vegetation patterns, a natural phenomenon observed worldwide,
are typically driven by spatially distributed feedback.
However, the spatial colonization mechanisms of clonal plants,
driven by the growth of a rhizome, are usually not considered
in prototypical models. Here we propose a general equation for
the vegetation density that includes all main clonal-growth
features as well as the essential ingredients leading to
spatial self-organization. This generic model reproduces the
phase diagram of a fully detailed model of clonal growth. The
relation of each term of the model with the mechanisms of
clonal growth is discussed.

\end{abstract}

\maketitle

\section{INTRODUCTION}
The spatial distribution of vegetation is a key factor in
ecosystems functionality as it may completely reorganize the
pathways of energy and resources \cite{MeronBook}. Besides the
simple homogeneous coverage, and disordered configurations,
several types of inhomogeneous vegetation distributions have
been reported, ranging from isolated gaps, scattered gaps
arranged in a more or less regular lattice, stripes or
labyrinthine patterns, patches arranged in a regular lattice, or
isolated patches \cite{MeronBook,gowda2015}. Although there are
a variety of mechanisms responsible for creating and
maintaining the spatial inhomogeneities, they are always
associated to feedback across space
\cite{Rietkerk2008,Getzin16} from which similar patterns arise
in completely different environments. Even the sequence in
which the different patterns appear when changing a control
parameter is often the same \cite{gowda2015}, which gives a
universal character to the phenomenon of vegetation pattern
formation.

Most studies of vegetation patterns consider arid environments,
where water is the limiting factor. Different
approaches have been considered, ranging from simple models
describing vegetation only
\cite{lefever1997origin,Lejeune1999,Lejeune2002}, to more
sophisticated ones accounting for vegetation and water
\cite{Getzin16,FernandezOto2019}. Long-range competitive
mechanisms are usually the reason behind pattern formation,
sometimes mediated by a diffusive external agent such as water
or described effectively by an interaction kernel. Thus, the
selected wavelength of the pattern is the result of the
interaction of two spatial scales, the range of competition, and
the spatial scale given by the diffusion of vegetation.

In these models, plant competition for water is the basic
factor introducing destabilizing feedback. On the other hand,
vegetation propagation is usually assumed to occur by seed
dispersal. However, a recent study of pattern formation in
underwater meadows of seagrasses \cite{ruiz2017fairy} was
clearly outside the domain of applicability of these two
standard hypothesis. First, although the mechanism for plant
interaction was not uniquely identified, competition for water
can not be a relevant interaction mechanism for marine plants.
Second, the main mode of reproduction of seagrasses is not seed
production, but clonal growth. Clonal plants (examples include
most grasses and seagrasses) reproduce by originating new
plants from a rhizome which grows horizontally. The rhizome, in
turn, can branch, creating a new rhizome propagating in a
different direction. Altogether, clonal plants can expand
without the need of producing seeds or spores, although most
species alternate clonal and sexual modes of reproduction under
some circumstances.

A numerical model to describe meadows of clonal plants, the ABD
model, was proposed in Ref. \cite{ruiz2017fairy} and it successfully
reproduced the observed patterns in seagrass meadows. However,
the model was highly complex due to the need to account for the
direction of growth of the rhizomes. In this paper, we
propose a single partial differential equation for the
vegetation density which reproduces qualitatively the patterns and
dynamics of clonal-plant meadows. We first derive the model
heuristically from the main mechanisms of growth and symmetry
properties. In this way, the model is a generic one, which could
be, in principle, applied to any instance of clonal growth. The
specific mechanisms of feedback and competition would only
enter through the particular values of the model parameters.
Then, we also derive the equation from the fully detailed ABD
model under certain approximations.
This allows us to relate the underlying growth mechanisms with
the different terms in the simplified description.

\section{THE CLONAL-GROWTH MODEL}
In the following we propose a heuristic large-scale model
(meadow or landscape scales) for a clonal-growth vegetation
density $n(\vr,t)$. This quantity gives the biomass, or number
of shoots per unit area, at location $\vr$ in a two-dimensional
location in the meadow. The model takes the form of a single
partial differential equation for $n(\vr,t)$ and is derived
under four general considerations: First, the homogeneous
unpopulated solution [i.e. $n(\vr,t)=0$, representing bare
soil] should always be a solution of the equation
for any parameter values (we do not consider the
possibility of plant immigration from outside the meadow).
Second, as usual when deriving large-scale equations which are
supposed to represent generic pattern formation processes
\cite{Cross1993}, we include in the equation only low-order
polynomial dependencies in the density and in its lowest order
gradients. Third, despite individual rhizomes grow in different
directions, at large scales, and not close to vegetation
borders, we should have all growth directions locally
represented and then the equation for the total density of
plants growing in all directions, $n(\vr,t)$, should be
rotationally invariant in the plane. And fourth, $n(\vr,t)$ can
never be negative.

Taking into account these requirements, a general partial
differential equation that contains in its right-hand side
polynomial terms up to third order in $n$ and up to fourth order in
gradients is:
\begin{eqnarray}
\partial_t n &=& -\omega n + an ^2 -b n ^3  + \epsilon \nabla^2 n   \nonumber \\
&+& \alpha ( \nabla ^2 n ) n + \delta \norm*{\vec{\nabla} n}^2 + \beta (\nabla^4 n)n \ .
\label{original_eq}
\end{eqnarray}
The absence of terms independent of $n$ implements our first
requirement. Also, only the rotationally invariant terms
containing gradients are present in Eq.~(\ref{original_eq}). A
term containing $\nabla^4 n$ could in principle be added to Eq.
(\ref{original_eq}), but it does not add any qualitatively
new behavior with respect to the terms already present. We
neither include terms such as $\norm*{\nabla^2 n}^2$, nor
others of higher order, since Eq.~(\ref{original_eq}) already explains
the relevant phenomenology, and it will be later derived in this
form from the ABD model.

In Eq.~(\ref{original_eq}), $\omega$ is readily interpreted as
the local net death rate in the linear regime, i.e. in the
absence of plant interactions. Negative values of $\omega$ would
indicate local linear net growth. In clonal plants, most of the
growth occurs by rhizome elongation which, in contrast to
plant death, is not a strictly local process. Local growth
occurs only when there is rhizome branching, and then the local
net death rate $\omega$ should be the difference between a linear
death rate $\omega_{d0}$ and a rhizome branching rate $\omega_b$:
$\omega=\omega_{d0}-\omega_b$. $a$ and $b$ account for local facilitative and
competitive interactions. The signs are chosen so that $a>0$
corresponds to a facilitative interaction (decrease of death
rate with density). To have a finite maximum
homogeneous density we need $b>0$, which models plant
competition (increase of death rate with density).

Equations similar to Eq. (\ref{original_eq}) have been derived from
models of dryland vegetation under competition for
water \cite{Lejeune1999,Lejeune2002,FernandezOto2019} or in
more general contexts involving species competition
\cite{Paulau2014}. In these contexts, the terms proportional to $\alpha$ and
$\beta$ are recognized as interaction terms that respect the
requirements of existence of a bare-soil solution and
positivity of the density. They are an effective way
to include, to lowest order in gradients, interactions between
distant plants mediated by water or any other long-range
competitive process
\cite{Lejeune1999,Lejeune2002,FernandezOto2019,Paulau2014}. In
these models, the diffusive term $\epsilon \nabla^2 n$ accounts
for plant propagation by seed dispersion. The new term here,
absent in previous works, is $\delta \norm*{\vec{\nabla} n}^2$.
When $\delta>0$, it always produces an increase in density at
any place where there is a non-zero gradient. This is the
effect that clonal growth by rhizome elongation would produce,
so that we interpret this term as the distinct signature of
clonal reproduction. Derivation of Eq. (\ref{original_eq}) from
the detailed model will confirm this.

\section{DERIVATION OF THE MODEL}
We next attempt a systematic derivation of
Eq.~(\ref{original_eq}) from the ABD full model
\cite{ruiz2017fairy} under certain approximations. This will
allow us to express the parameters in Eq. (\ref{original_eq}) in
terms of biologically relevant ones.

Previous works have approximated the space occupation of clonal
plants from a simple random-walk process
\cite{routledge1990spatial} or through the definition of
discrete growth rules \cite{sintes2005nonlinear}. The latter
work identified three key ingredients. First, the rhizome of a
plant, whose tip is called the apex, grows horizontally at constant
velocity $\nu$, leaving behind new shoots separated by a
characteristic distance $\rho$. Second, the shoots and apices
have a lifetime depending on the environmental conditions which
translates into a mortality rate $\omega_d$. Finally, the
rhizomes can generate new branches growing in other directions
separated by a characteristic angle $\phi_b$ from the initial
one. Branching happens with a rate $\omega_b$. The ABD model is
an implementation at the landscape level and in terms of
population densities of the above mechanisms. The time
evolution of the spatial and angular density of apices
$n_a(\vec{r},\phi,t)$, where $\phi$ is the angle of the growth
direction, and of shoots $n_s(\vec{r},t)$, is ruled by
\begin{eqnarray}
\label{E1}
\partial_t n_a(\vec{r},\phi,t) &&= -\omega_d[n_t] n_a(\vec{r},\phi,t)
-\vec{v}(\phi)\cdot\vec{\nabla} n_a(\vec{r},\phi,t) \nonumber \\
&+& \frac{\omega_b}{2}\left(n_a(\vec{r},\phi+\phi_b,t)
+ n_a(\vec{r},\phi-\phi_b,t)\right) \\
\label{E2}
\partial_t n_s(\vec{r},t) =&-&\omega_d[n_t] n_s(\vec{r},t) + \frac{\nu}{\rho}\int_0^{2\pi} n_a(\vec{r},\phi,t) d\phi,
\end{eqnarray}
where the growth-velocity vector is
$\vec{v}=\nu(\cos\phi,\sin\phi)$ and
$\vec{\nabla}=(\partial_x,\partial_y)$. The first term on the
right-hand-side of Eq.~(\ref{E1}) accounts for the mortality of apices,
the second is an advection term that displaces apices in the
direction of growth $\phi$ due to the elongation of the
rhizome. The third corresponds to the branching process where
the density of apices in adjacent directions $\phi\pm\phi_b$
contribute to increase the density of apices growing with the
direction $\phi$. The first term in Eq. (\ref{E2}) corresponds
to shoots death (same mortality rate is assumed for shoots and
apices) and the second contribution accounts for shoots left
behind the apices while the rhizomes grow in all directions.

The death rate of the plant $\omega_d$ depends on the total
density $n_t(\vec{r},t) = n_s(\vec{r},t) + N_a(\vec{r},t)$,
where $N_a(\vec{r},t) = \int_0^{2\pi} n_a(\vec{r},\phi,t)
d\phi$ is the density of apices growing in all directions, as:
\begin{equation}
\label{E3}
\omega_d[n_t(\vec{r},t)]= \omega_{d0} + B n^2_t +\int \int
\mathcal{K}(\vec{r}-\vec{r}')(1-e^{-a_e n_t(\vec{r}')}) d\vec{r}'.
\end{equation}
The first contribution accounts for the death rate of a single
isolated shoot or apex. The second is a saturating term
accounting for the environmental carrying capacity. Finally,
the third one is an integral term which describes in a very
general way the interactions across space. The kernel
$\mathcal{K}$ includes both facilitative and competitive
interactions. It does not assume any specific interaction
mechanism, but encodes its strength and spatial scales.

For appropriate choices of the parameters and of the kernel
$\mathcal{K}$, this model can describe accurately growth and
pattern formation in seagrass meadows \cite{ruiz2017fairy}. One
of the particularities of the model is that it includes
explicitly the directions of growth. Although this provides a
complete description, it is highly demanding from a
computational point of view, as one has to deal with a
three-dimensional field for the apices ($n_a$ depends on
$\vec{r}$ and $\phi$) plus a two-dimensional field for the
shoots. One can eliminate the dependence on the angle and find
a single equation for the total density by introducing a number
of approximations detailed in the Appendix. The result is that
an approximate equation for the total density reads
\begin{equation}
\label{ABDAvec1}
  \partial_t n_t = (\omega_b -\omega_d[n_t])n_t-\nu \nabla\cdot\vec{a}' \ ,
\end{equation}
and
\begin{equation}
  \label{aprox3}
   \vec{a}'=-(c_0+c_1n_t)\vec{\nabla}n_t.
\end{equation}
$c_0$ and $c_1$ are approximately given by
\begin{eqnarray}
  c_0 &=& \frac{\nu}{2 (\omega_{d0,M}-\omega_b\cos\phi_b)}, \\
  \label{c0}
  c_1 &=& \frac{\nu}{2 n_{t,M}^*} \left(\frac{1}{\omega_b\cos\phi_b-\omega_{d0,M}}-\frac{1}{\omega_b\cos\phi_b-\omega_b}\right),\
  \label{c1}
\end{eqnarray}
being $n_{t,M}^*$ the homogeneous stationary value of density
at the Maxwell point ($\omega_{d0}=\omega_{d0,M}$), where a
front between the populated homogeneous solution and bare soil
does not move.

The term $\omega_b -\omega_d[n_t]$ in Eq.~(\ref{ABDAvec1}) can
be interpreted as a net growth rate at location $\vr$ depending
on the density in the surroundings (because of the integral
term in $\omega_d[n_t]$). The form of Eq.~(\ref{ABDAvec1}) also
indicates that $\vec{a}'$ is a flux of biomass arising from
propagation mechanisms. These were just clonal growth in the
original equations of the ABD model. Thus, the propagation
contribution to Eq.~(\ref{ABDAvec1}),
\begin{equation}
-\nu\nabla\cdot \vec{a}' =
\nu \left(c_0 \nabla^2 n_t + c_1 n_t \nabla^2 n_t + c_1 \norm*{\vec{\nabla}n_t}^2 \right) \
  \label{propagation}
\end{equation}
encodes the contribution from clonal growth. The first and the
second terms in Eq. (\ref{propagation}) have a functional form
already encountered in other models of vegetation dynamics,
where they accounted for, respectively, seed dispersion and
interactions. But here they arise from multidimensional rhizome
growth and branching and, as anticipated, the presence of the
new term $\norm*{\vec{\nabla}n_t}^2$ is a distinct signature of
this clonal mode of propagation.

To follow further toward the derivation of Eq.
(\ref{original_eq}), we approximate the exponential term
$(1-e^{-a_e n_t})$ in Eq. (\ref{E3}) to first order in $a_e n_t$,
and expand the resulting integral using a moment expansion
\citep{murray2002mathematical}. Then we obtain the following
expression for the nonlocal interacting terms in $\omega_d$:
\begin{equation}
\int \int
\mathcal{K}(\vec{r}-\vec{r}')(1-e^{-a_e n_t(\vec{r}')}) d\vec{r}'\simeq a_e\sum_{j=0}^\infty d_j \nabla^{2j}n_t(\vec{r},t),
\label{simply_int}
\end{equation}
where
\begin{eqnarray}
d_j &=& \frac{(-1)^{j}}{(2j)!} 2\pi J_0^{(2j)}(0)\int\limits^{\infty}_{0}r^{2j+1}\mathcal{K}(r)dr
\label{moments}
\end{eqnarray}
are the corresponding moments of the kernel and $J_0^{(2j)}(0)$
is the $2j$th derivative of the Bessel function $J_0$ of the
first kind evaluated at the origin. Considering terms only
until fourth order and replacing them, together with
Eq.~(\ref{propagation}), in Eq. (\ref{ABDAvec1}) we get exactly
Eq.~(\ref{original_eq}) with the identification of $n(\vr,t)$
with the total density $n_t(\vr,t)$, and
$\omega=\omega_{d0}-\omega_b$, $a=-a_e d_0$, $b=B$,
$\epsilon=\nu c_0$, $\delta=\nu c_1$, $\alpha=\nu c_1 - a_e
d_1$, and $\beta=-a_e d_2$.

We can now identify the contribution of the different
mechanisms to each term of Eq.~(\ref{original_eq}). As
anticipated, the difference between the mortality and branching
rates $\omega_{d0}$ and $\omega_b$ determines the net growth
$\omega$, but the rate $\omega_b$ appears also explicitly in
other coefficients. The parameters related to rhizome
propagation, $\nu$ and $\phi_b$, determine the value of the
diffusion and nonlinear diffusion constants $\epsilon$ and
$\alpha$, as well as the value of the coefficient $\delta$.
This illustrates the role the growth of the rhizome and the
branching play in colonizing empty space.

\section{VEGETATION PATTERNS}
\begin{figure}[t]
    \centering
    \includegraphics[width=0.95\columnwidth]{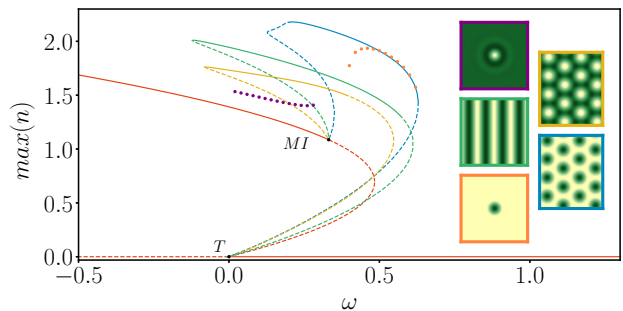}
    \caption{
Bifurcation diagram of the homogeneous
    steady states $n^*$ (red lines)
    and stationary patterns of Eq. (\ref{original_eq}). The maximum value of the
    density $n(\vec{r})$ is plotted as a function of mortality $\omega$.
    Solid (dashed) lines indicate stable (unstable) solutions. The 2$d$
    spatial distribution of each solution is shown in the corresponding inset: negative
    hexagons (yellow line), stripes (green line), and positive hexagons
    (blue line). Dots indicate the stable part of the branches
    of localized states: negative soliton (purple dots),
    and positive soliton (orange dots).
    Parameters: $a=1.39$, $b=1$, $\epsilon=1.15 \times 10^{-2}$, $\alpha=-1.78$,
    $\delta=1.03 \times 10^{-2}$, and $\beta=-1$.
    }
    \label{homo_pattern}
\end{figure}

We next analyze some predictions of the model given by Eq.
(\ref{original_eq}). In the following, by rescaling $n$
and $\vec{r}$, we set $b=1$ and $\beta=-1$
without loss of generality. Equation ~(\ref{original_eq}) has three
homogeneous steady states (HSSs): $n^{*}_0=0$ and
$n^{*}_{\pm}=(a\pm\sqrt{a^2-4\omega})/2$. The first one
corresponds to bare soil (unpopulated solution), while the
other two emerge from a saddle node bifurcation located at
$a^2-4\omega=0$, so that $n^{*}_\pm$, representing  homogeneous
populated solutions, exist only for sufficiently low
mortalities, $\omega<a^2/4$. Since $n$ is a density, only
positive values are considered in the following. Depending on
the sign of $a$, either $n^{*}_+$ or $n^{*}_-$ connect with
$n^{*}_0$ in a transcritical bifurcation at $\omega=0$. With
facilitative interactions, $a>0$, $n^{*}_+$ is stable against
homogeneous perturbations while $n^{*}_-$ is unstable and
connects with $n_0$ subcritically. As a result there is a
region of coexistence between the populated and the unpopulated
states. On the other hand, if $a<0$ only $n^{*}_+$ takes
positive values and the transition is supercritical. Here we
focus in the case $a>0$ as in Ref. \cite{ruiz2017fairy}. Figure
\ref{homo_pattern} shows an example of the bifurcation diagram
of the homogeneous solutions as a function of the mortality
$\omega$ in this subcritical case. In this and in the next
figures, parameters used in our simplified model are determined
from the ones appropriate for the seagrass \emph{Posidonia
oceanica} in the ABD model \cite{ruiz2017fairy}.

\begin{figure}[ht]
    \centering
    \includegraphics[width=0.95\columnwidth]{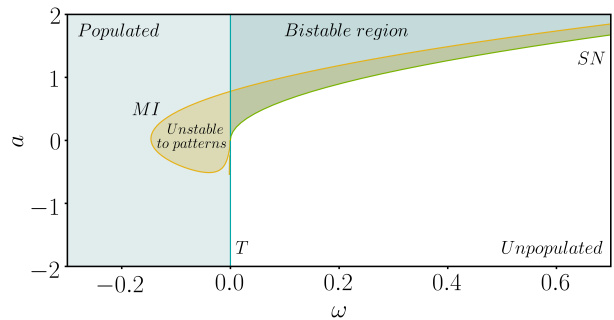}
    \caption{Phase diagram of the homogeneous steady states of
    Eq.~(\ref{original_eq}). Blue line at $\omega=0$ signals the
    transcritical bifurcation of the zero solutions. For $a<0$,
    this bifurcation is supercritical and the populated solutions
    $n^*_+$ exist only for $\omega<0$. For $a>0$, the bifurcation
    is subcritical and the region of existence of $n^*_+$ extends
    to positive values of $\omega$ until the saddle-node bifurcation
    (SN) indicated by the green line. In the white region, the only
    possible solution is bare soil. In the light blue region, the
    only stable HSS is $n^*_+$, while in the dark blue region $n^*_+$
    stably coexist with bare soil. The yellow line signals the
    modulation instability of $n^*_+$. In the yellow regions, $n^*_+$
    is unstable to patterns. The region of existence of patterns
    extends beyond the yellow region. Parameters as in Fig. \ref{homo_pattern}.}
     \label{mi_color}
\end{figure}

Figure \ref{mi_color} shows the location of different bifurcation
and stability domains in the two-parameter space
($\omega$,$a$), as identified from a linear stability analysis
of the HSS against perturbations of the form $e^{\lambda t +
i\vec{q}\cdot\vec{x}}$. It reveals that the unpopulated
solution $n_0^{*}$ is unstable for $\omega < 0$, leading to a
homogeneous growth of the density until the solution $n_+^{*}$
is reached. This branch $n_+^{*}$ is stable against homogeneous
perturbations $q=0$, where $q=\|\vec{q}\|$. However, for
$\omega$ above a certain threshold $\omega_c$, it becomes
unstable against modulations with a given critical wave number
$q_c$, what is known as a modulation (or Turing) instability
(MI). For values of $\omega>\omega_c$, the HSS $n_+^{*}$
remains unstable until the saddle node bifurcation (see
Figs.~\ref{homo_pattern} and \ref{mi_color}). In the
supercritical case, $a<0$, the homogeneous state becomes
unstable at the MI and the region of instability persists until
a second MI close to the transcritical bifurcation at
$\omega=0$. The phase diagram in Fig.~\ref{mi_color} reproduces
all the qualitative features present in the ABD model (compare
with Fig. S3 in Ref. \cite{ruiz2017fairy}). Beyond the qualitative
agreement, the shape and velocity of a front between the
populated and unpopulated solutions, as well as the position in
parameter space of the MI in the ABD model, can be
quantitatively predicted by the reduced model if the full
nonlocal term is kept, but this quantitative accuracy is
partially lost when the expansion of the integral term is
truncated linearly in $n_t$ and to fourth order in the
gradients, which is the roughest approximation in our
derivation.

The nonlinear regimes of the dynamics can not be inferred just
by the linear stability analysis. We use
numerical continuation techniques to follow the stable and
unstable parts of the branches of different patterns. Localized
structures are obtained using numerical simulations, capturing
only the stable part of the branch. Figure \ref{homo_pattern}
shows the different patterns observed for different values of
the growth rate $\omega$. Although the value of the parameters
for which the different patterns appear does not precisely
correspond with the ones in the ABD model \cite{ruiz2017fairy},
the type and sequence of patterns when the mortality is
increased are the same, as expected from the general theory
\cite{gowda2015}.

\section{EFFECTS OF CLONAL GROWTH}

\begin{figure}
    \centering
    \includegraphics[width=0.95\columnwidth]{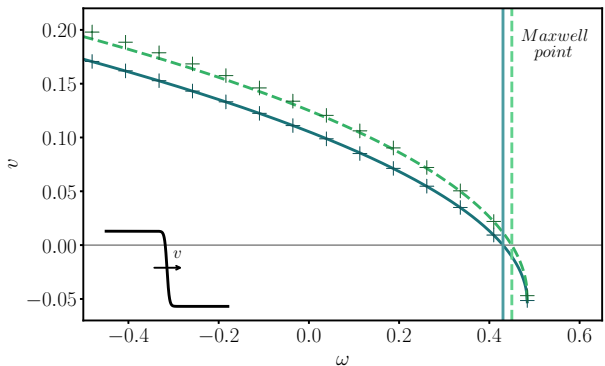}
    \caption{Velocity of a front between bare-soil
    and the homogeneous populated solution (in the absence of the MI) as a function of the mortality $\omega$.
    Crosses are the results of numerical simulations. The curve in blue (solid) corresponds to the analytical solution
    of Eq.~(\ref{original_eq}) for $\alpha=\beta=\delta=0$ (see Appendix).
    The curve in green (dashed) is the velocity computed using the
    perturbative method detailed in the Appendix for $\alpha=\beta=0$ and
    to first order in $\delta$ for $\delta=1.03 \times 10^{-2}$. Other parameters as in Fig. \ref{homo_pattern}. The solid (dashed) vertical lines indicate the position of the Maxwell point for $\delta=0$ ($\delta=1.03 \times 10^{-2}$).
    }
     \label{velocity}
\end{figure}

The term $\delta \norm*{\vec{\nabla}n_t}^2$,
which originates from the distinctive
mechanism of clonal growth, affects mainly the dynamics of
fronts. This term contributes always to the velocity of a front
in the direction from higher to lower densities.
In the absence of the MI, using the perturbative method detailed in the Appendix, it is possible to calculate the contribution of this term to the
velocity of a front between bare soil and
the homogeneous populated state. The MI is not present when $\alpha=\beta=0$, which corresponds to the absence of long-range interactions, and the
growth rate of perturbations with wave number $q$ goes as $-\epsilon q^2$, where the maximum growth rate corresponds to the homogeneous mode. In this case, as it can be seen in Fig.
\ref{velocity}, the term with coefficient $\delta$ favors the expansion (i.e., increases
the velocity) of the populated solution over the zero state,
which we interpret as the result of the elongation of the
rhizomes outward of the meadows. As a consequence, the position of
the Maxwell point where the front has zero velocity moves to
higher mortalities as compared to the same equation with
$\delta=0$ \cite{Alvarez17}, i.e. clonal-growth plants can
colonize a new empty space under unfavorable conditions more
efficiently than if the propagation is driven by seed dispersal
(diffusion) only.

\begin{figure}
    \centering
    \includegraphics[width=0.92\columnwidth]{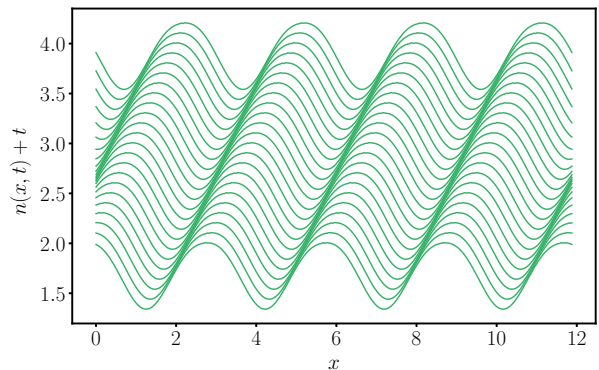}
    \caption{Space-time representation of a
    pattern of stripes traveling toward the right for
    $\omega=0$, $\alpha=-8$ and
    $\delta=15$. Other parameters as in Fig. \ref{homo_pattern}. Densities at successive times are displaced vertically. For the same parameter values
    a parity-symmetric solution exists that travels toward the left.
    }
     \label{travpat}
\end{figure}

Further increasing $\delta$ can have important effects on the
dynamics of patterns. For instance, for large values of
$\delta$ traveling patterns are observed as
shown in Fig. \ref{travpat}. The figure shows a periodic
pattern traveling to the right, whereas another solution (with
shape related by the $x \rightarrow -x$ parity transformation)
with velocity toward the left also exists. The velocity of the
patterns grows, increasing $\delta$. Such parity-breaking
bifurcation from steady to moving patterns is actually observed
in the ABD model for low branching angles $\phi_b$. This is
consistent with the relation between $\phi_b$ and $\delta$
derived in this work [Eq. (\ref{c1})]. Thus,
low branching angles, which are common for different species,
can lead to bigger values of $\delta$, which can trigger this
instability. The detailed analysis of the transition to moving
patterns as well as other dynamical regimes of Eq.
(\ref{original_eq}) will be studied elsewhere.

\section{CONCLUSIONS}

Summarizing, we have proposed a simple model to describe the
growth and dynamics of clonal-plant meadows. We have also
derived the equation from a realistic model, providing
analytical expressions for the effective parameters as a
function of the biologically relevant parameters of the full
model. The reduced model provides a qualitative description of
clonal-growth plants, reproducing all the stationary spatial
distributions and dynamical regimes. Moreover, beyond a
qualitative description, accurate quantitative results can be
obtained depending on the level of approximation of the
non-local interacting terms. We expect this simple model,
applicable to a wide variety of clonal plants, to allow deeper
theoretical studies on the dynamics of clonal growth not
tractable using a fully detailed model.

\acknowledgments We acknowledge financial support from FEDER/Ministerio de Ciencia, Innovaci\'on y Universidades – Agencia Estatal de Investigaci\'on\// SuMaEco project (RTI2018-095441-B-C22) and the Mar\'ia de Maeztu Program for Units of Excellence in R\&D (No. MDM-2017-0711). D.R.-R. also acknowledges the fellowship No. BES-2016-076264 under the FPI program of MINECO, Spain.

\section*{Appendix}
\renewcommand{\theequation}{A\arabic{equation}}
\setcounter{equation}{0}  
\renewcommand{\thesubsection}{\arabic{subsection}}
\setcounter{subsection}{0}  

In this Appendix we give details on the derivation of our
simplified equation from the ABD model, and on the calculation
of the velocity of a front using perturbative methods.

\subsection{Derivation of the clonal-growth equation}

We derive here our simplified equation for the spatial
distribution of the density $n(\vr,t)$ of clonal-growth
vegetation:
\begin{eqnarray}
\partial_t n &=&
- \omega n + an ^2 -b n ^3  + \epsilon \nabla^2 n  \nonumber \\
&+& \alpha ( \nabla ^2 n ) n + \delta \norm*{\vec{\nabla} n}^2 + \beta (\nabla^4 n)n \ ,
\label{SM:original_eq}
\end{eqnarray}
starting from the more detailed and biologically motivated ABD
model \cite{ruiz2017fairy}. This will allow us to give a one to
one correspondence of the parameters in Eq. (\ref{SM:original_eq})
with biologically relevant parameters.

The ABD model, introduced in Ref. \cite{ruiz2017fairy}, describes
the time evolution of the spatial and angular density of apices
$n_a(\vec{r},\phi,t)$, where $\phi$ is the angle of the growth
direction, and shoots $n_s(\vec{r},t)$, in a meadow,
\begin{eqnarray}
\label{SM:E1}
\partial_t n_a(\vec{r},\phi,t)&& = -\omega_d[n_t] n_a(\vec{r},\phi,t)
-\vec{v}(\phi)\cdot\vec{\nabla} n_a(\vec{r},\phi,t) \nonumber \\
+&\frac{\omega_b}{2}& \left(n_a(\vec{r},\phi+\phi_b,t)
+ n_a(\vec{r},\phi-\phi_b,t)\right), \\
\label{SM:E2}
\partial_t n_s(\vec{r},t) =&-&\omega_d[n_t] n_s(\vec{r},t) + \frac{\nu}{\rho}\int_0^{2\pi} n_a(\vec{r},\phi,t) d\phi,
\end{eqnarray}
where the growth-velocity vector is
$\vec{v}=\nu(\cos\phi,\sin\phi)$ and
$\vec{\nabla}=(\partial_x,\partial_y)$.

The death rate of the plant $\omega_d[n_t(\vr,t)]$ depends on
the total density $n_t(\vec{r},t) = n_s(\vec{r},t) +
N_a(\vec{r},t)$, where $N_a(\vec{r},t) = \int_0^{2\pi}
n_a(\vec{r},\phi,t) d\phi$ is the total density of apices
growing in all directions, as
\begin{equation}
\label{SM:E3}
\omega_d[n_t(\vec{r},t)]= \omega_{d0} + B n^2_t +\iint\limits_{-\infty}^{+\infty}
\mathcal{K}(\vec{r}-\vec{r}')(1-e^{-a_e n_t(\vec{r}')}) d\vec{r}',
\end{equation}
where the kernel $\mathcal{K}$ is the difference of two
normalized Gaussians $\mathcal{G}$ of different strengths
$\kappa$ and $\mu$, and widths $\sigma_\kappa$ and
$\sigma_\mu$:
\begin{equation}
\label{E8}
\mathcal{K}(\vec{r})=\kappa
\mathcal{G}(\sigma_\kappa,\vec{r})-\mu\mathcal{G}(\sigma_\mu,\vec{r}) \ .
\end{equation}
We first obtain a relationship for the steady homogeneous
solutions, $n_s(\vec{r},t)=n_s^*$ and
$n_a(\vec{r},\phi,t)=n_a^*$, from which $N_a^*=2\pi n_a^*$.
From Eq. (\ref{SM:E1}), we obtain that
$(-\omega_d[n_t]+\omega_b) n_a^*=0$, which, for a populated
solution, implies equality between branching and death rates,
$\omega_b=\omega_d[n_t]$. Using this into the steady state of
Eq. (\ref{SM:E2}), we get $(\nu/\rho) N_a^* - w_b n_s^*=0$, or
$n_s^*=(\nu/\rho\omega_b) N_a^*$. Inserting in
$n_t^*=n_s^*+N_a^*$, we find the following relationship between
the total-apex density and the total plant density:
\begin{equation}
N^*_a=  \frac{\rho\omega_b}{\nu +\rho\omega_b}n_t^* \equiv \eta n_t^*.
\label{SM:ASrel}
\end{equation}

To eliminate the angular dependence in Eqs.
(\ref{SM:E1}) and (\ref{SM:E2}), and find a single equation for
the total density, we first write the density of apices as a
Fourier series in the angle $\phi \in [0,2\pi]$:
\begin{align}
  & n_a(\vec{r},\phi,t) = \frac{a_0 (\vec{r},t)}{2} \nonumber \\
&+ \sum_{m=1}^{\infty} \left( a_m (\vec{r},t) \cos (m\phi) +
b_m (\vec{r},t) \sin (m\phi) \right). \
  \label{SM:fouriern_a}
\end{align}
Note that $\pi a_0 (\vec{r},t)= N_a (\vec{r},t)$. Using Eqs.
(\ref{SM:E1}) and (\ref{SM:fouriern_a}), we find the evolution
equations for all the amplitudes $a_m$, $b_m$. This gives an
infinite hierarchy of coupled equations such that modes $m$ are
coupled to modes $m+1$. From numerical simulations and the
linear stability analysis one can check, however, that modes
with $m>1$ are not contributing substantially to the dynamics.
Therefore these terms can be neglected. As a second
approximation we assume that the relation in Eq. (\ref{SM:ASrel}),
obtained for the populated HSS, is also
valid for all $\vec{r}$ and $t$ in any heterogeneous spatial
distribution. We have checked the accuracy of this
approximation using numerical simulation. The maximum error in
a stationary pattern is less than $10\%$ of the total
density of apices.

With these approximations Eqs. (\ref{SM:E1}) and (\ref{SM:E2})
can be simplified to three equations (for $n_t$, $a_1$, and
$b_1$). Defining $\vec{a} \equiv (a_1,b_1)$,
$A=\norm*{\vec{a}}$ and $\theta=\arctan(b_1/a_1)$, they can be
written as
\begin{align}
\label{SM:ABDAvec1}
\partial_t n_t &= (\omega_b -\omega_d[n_t])n_t-\frac{\nu\pi}{\eta} \nabla\cdot\vec{a}   \\
\label{SM:ABDApol1}
\partial_t A &= (\omega_b\cos\phi_b -\omega_d[n_t])A-\frac{\nu\eta}{2\pi}\norm*{\vec{\nabla}n_t}\cos(\theta-\gamma) \\
\label{SM:ABDApol2}
\partial_t \theta &= \frac{\nu\eta}{2\pi} \frac{\norm*{\vec{\nabla}n_t}}{A}\sin(\theta-\gamma) \ ,
\end{align}
where $\gamma(\vr,t)=\arctan(\partial_y n_t/\partial_x n_t)$ is
the angle that the gradient of the total density forms with the
$x$ axis. From Eq. (\ref{SM:ABDApol2}), the evolution of
the system will drive the angle $\theta$ to the stable fixed
point $\theta=\gamma+\pi$. This means that, in the long run,
the density gradient and the vector $\vec{a}$ have opposite
directions: $\vec{a} = - C \vec{\nabla}n_t$, or $A=
C\norm*{\vec{\nabla}n_t}$, with $C$ a positive constant.
Introducing this result in the Fourier series given by Eq.
(\ref{SM:fouriern_a}) truncated at $m=1$, the density of apices
takes the following form:
\begin{equation}
  n_a(\vec{r},\phi,t) = \frac{N_a(\vec{r},t)}{2\pi} + C\norm*{\vec{\nabla}n_t(\vec{r},t)}\cos(\phi-\gamma(\vec{r},t)-\pi) \ .
\end{equation}

\begin{figure}[t]
    \centering
    \includegraphics[width=\columnwidth]{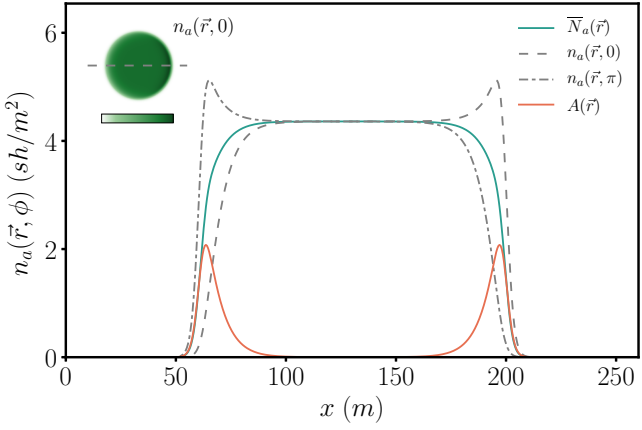}
    \caption{Densities (in shoots per square meter, sh/m$^2$) of apices of a circular domain invading the
    unpopulated solution for the ABD model with parameters appropriate for the seagrass
    \emph{Posidonia oceanica} but without nonlocal interactions \cite{ruiz2017fairy}: $\omega_b=0.06$ year$^{-1}$,
    $\omega_{d0}=0.038$ year$^{-1}$ , $\nu=6.11$ cm/year, $\rho=2.87$ cm,
    $\phi_b=45^\circ$, $b=1.25$ cm$^4$year$^{-1}$, $\kappa=0.048$ year$^{-1}$,
    $\sigma_\kappa=0$ cm, $a=27.38$ cm$^2$, $\sigma_\mu=0$cm, $\mu=\omega_{d0}$.
    The inset in green represents the density of apices growing to the right ($\phi=0$),
    and the dashed line indicates the cut shown in the main plot. The blue line
    represents the mean angular density of apices ($\bar{N}_a = N_a/2\pi$).
    The dashed (dot-dashed) line shows the density of apices growing to the right (left).
    The red line shows the amplitude $A$ of the first mode ($m=1$) in Eq.~(\ref{SM:fouriern_a}).
    }
    \label{SM:Apicesdist}
\end{figure}

In regions where the total density is homogeneous,
$\vec{\nabla}n_t$ is zero and  the densities of apices growing
in all directions are equal and proportional to the total
density of apices $N_a$. On the other hand, in regions where
$\vec{\nabla}n_t\neq0$, there is an angular modulation of the
density. For example, considering a circular patch, those
apices growing outward (normal to the vegetation border) are
denser while those growing inward are depleted, as can be
seen in Fig. \ref{SM:Apicesdist} from numerical simulations. The same effect can be observed for different spatial distributions as shown in Ref. \cite{Ruiz-Reynes2019}.
The value of $C$ is related to the amplitude of the modulation
at the borders of the patch, and in general it will be a
complicated function of $n_t$ and its derivatives. Considering
only the leading terms of a power expansion of
$C(n_t,\vec{\nabla}n_t,...)$ we take:
\begin{equation}
  \label{SM:aprox3}
   \vec{a}=\frac{\eta}{\pi}\vec{a}'=-\frac{\eta}{\pi}(c_0+c_1n_t)\vec{\nabla}n_t \ ,
\end{equation}

where $c_0$ and $c_1$ are constants to be determined.
Introducing Eq. (\ref{SM:aprox3}) in Eq. (\ref{SM:ABDAvec1}), one
obtains a closed equation for the total density $n_t$:
\begin{equation}
  \partial_t n_t = (\omega_b-\omega_d[n_t])n_t +
  \nu(c_0 \nabla^2 n_t + c_1 \norm*{\vec{\nabla}n_t}^2 + c_1 n_t \nabla^2 n_t) \ .
  \label{SM:laecuacion}
\end{equation}
This equation is used in the main text, together with an
expansion of the integral in $\omega_d[n_t]$ and identifying
$n$ with the total plant density $n_t$, to establish the
simplified model Eq. (\ref{SM:original_eq}).

Parameters $c_0$ and $c_1$ have not been determined so far. It
is possible to determine their value close to the Maxwell point
where a front between the populated HSS
and bare soil is at rest. Assuming such a stationary front and
introducing Eq. (\ref{SM:aprox3}) in Eq. (\ref{SM:ABDApol1}),
recalling that $\theta=\gamma+\pi$, one obtains the following
expression:
\begin{equation}
  \bigg\{(\omega_b \cos \phi_b-\omega_d[n_t])(c_0+c_1n_t)+\frac{\nu}{2}\bigg\}\frac{\eta}{\pi}\norm*{\vec{\nabla}n_t}=0.
  \label{SM:conditionc0c1}
\end{equation}
\textit{A priori} the front profile is not known, however, we know it
connects with the populated solution on one side and with the
unpopulated solution on the other, so we can write $n_t= n^*_t
+\varepsilon e^{\lambda x}$, where $n_t^*$ is zero for the
unpopulated solution at one side and takes the value of the
stationary homogeneous density $n_{t,M}^*$ for the populated
solution at the Maxwell point ($\omega_{d0}=\omega_{d0,M}$) at
the other. Here $\lambda$ is the spatial eigenvalue of each
solution. Introducing these expressions in Eq.
(\ref{SM:conditionc0c1}), at the lowest order in $\varepsilon$
we obtain
\begin{align}
  c_0 &= \frac{\nu}{2 (\omega_{d0,m}-\omega_b\cos\phi_b)}, \\
  c_1 &= \frac{\nu}{2 n_{t}^*} \left(\frac{1}{\omega_b\cos\phi_b-\omega_{d0,M}}-\frac{1}{\omega_b\cos\phi_b-\omega_b}\right).
\end{align}
For the unpopulated solution $n_t^*=0$ and the populated one
$n_t^*=n_{t,M}^*$.

The last approximation on the derivation consists of taking the
moment expansion of the integral term in Eq. (\ref{SM:E3}). We can
approximate the exponential term at first order and use the
convolution theorem to write:
\begin{align}
 \iint\limits_{-\infty}^{+\infty} \mathcal{K}(\vec{r}-\vec{r}')(1-e^{-a_e n_t(\vec{r}',t)}) d\vec{r}'  \nonumber \\
 \simeq a_e  \iint\limits_{-\infty}^{+\infty} \mathcal{K}(\vec{r}-\vec{r}') n_t(\vec{r}',t) d\vec{r}' \nonumber \\
 =\frac{a_e}{(2\pi)^2} \iint\limits_{-\infty}^{+\infty}  e^{i\vec{q}\cdot\vec{r}} \tilde{\mathcal{K}}(\vec{q}) \tilde{n}_t(\vec{q},t) d\vec{q},
\label{SM:simply_int1}
\end{align}
where $\tilde{\mathcal{K}}(\vec{q})$ and $\tilde{n}_t(\vec{q})$
are the Fourier transforms of the kernel and the total density.
Due to the symmetry of the kernel, $\tilde{\mathcal{K}}(\vec{q})
= \tilde{\mathcal{K}}(q)$, where $q= \|\vec{q}\|$.  We use a
Taylor expansion to write the kernel in terms of its
derivatives evaluated at the origin. Only even derivatives are
non-zero:
\begin{align}
\frac{a_e}{(2\pi)^2}\iint\limits_{-\infty}^{+\infty}  e^{i\vec{q}\cdot\vec{r}} \sum_{j=0}^\infty \frac{1}{j!}\frac{d^j \tilde{\mathcal{K}}(q)}{dq^j} \bigg|_{q=0}q^j \tilde{n}_t(\vec{q},t) d\vec{q}
\nonumber \\
=\frac{a_e}{(2\pi)^2} \sum_{j=0}^\infty \frac{1}{(2j)!}\frac{d^{2j} \tilde{\mathcal{K}}(q)}{dq^{2j}} \bigg|_{q=0} \iint\limits_{-\infty}^{+\infty}  e^{i\vec{q}\cdot\vec{r}} q^{2j} \tilde{n}_t(\vec{q},t) d\vec{q}
\nonumber \\
=a_e\sum_{j=0}^\infty d_j \nabla^{2j}n_t(\vec{r},t),
\label{SM:simply_int2}
\end{align}
where
\begin{eqnarray}
d_j = \frac{(-1)^{j}}{(2j)!} \frac{d^{2j}\tilde{\mathcal{K}}(q)}{dq^{2j}}\bigg|_{q=0}.
\label{SM:moments1}
\end{eqnarray}
The coefficients $d_j$ can be computed from Eq.
(\ref{SM:moments1}) or in terms of the moments of the kernel in
radial coordinates:

\begin{eqnarray}
d_j &=& \frac{(-1)^{j}}{(2j)!} \frac{d^{2j}\tilde{\mathcal{K}}(q)}{dq^{2j}}\bigg|_{q=0}
= \frac{(-1)^{j}}{(2j)!} \frac{d^{2j}}{dq^{2j}} \iint\limits_{-\infty}^{+\infty} e^{-i\vec{q}\cdot\vec{r}} \mathcal{K}(\vec{r}) d\vec{r} \bigg|_{q=0}\nonumber\\
&=& \frac{(-1)^{j}}{(2j)!} \frac{d^{2j}}{dq^{2j}} \int\limits_{0}^{\infty} \mathcal{K}(r) \int\limits_{0}^{2\pi} e^{-iqr \cos(\theta-\phi)} d\theta r dr \bigg|_{q=0}\nonumber\\
&=& \frac{(-1)^{j}}{(2j)!} \frac{d^{2j}}{dq^{2j}} \int\limits_{0}^{\infty} \mathcal{K}(r) 2\pi J_0(qr) r dr \bigg|_{q=0} \nonumber \\
&=& \frac{(-1)^{j}}{(2j)!} 2\pi J_0^{(2j)}(0)\int\limits^{\infty}_{0}r^{2j+1}\mathcal{K}(r)dr,
\label{SM:moments2}
\end{eqnarray}
where $J_0^{(2j)}(0)$ is the $2j$th derivative of the Bessel
function $J_0$ of the first kind evaluated at the origin.

\subsection{Calculation of the velocity of a front using a perturbative
method}

We consider Eq. (\ref{SM:E1}) in the case of $\alpha=\beta=0$. Thus,
the dispersion relation changes with the wave number $q$ as $-\epsilon q^2$, the maximum growth rate corresponds to $q=0$ and the modulational instability is absent. In this case, solutions consisting of an unpopulated and a populated homogeneous
solutions separated by a front that moves at constant velocity
$v$ exist, so that one can rewrite the equation in one
dimension as a time-independent one in the comoving reference
frame $x\rightarrow x-vt$. Positive velocity indicates motion
toward larger $x$ values. We want to analyze perturbatively
the effect of the presence of a small parameter $\delta$ on the
motion of the front, similarly to the approach in Ref. \cite{Lober2012}. To this end we write $\delta\rightarrow
\xi \delta$, with $\xi$ small, to emphasize that $\delta$ is
small. At the end of the calculation we can set back $\xi=1$.
The steady equation in the comoving frame reads:
\begin{equation}
0 = -\omega n + an ^2 -b n ^3  + \epsilon \partial_x^2 n + v\partial_x n
 + \xi \delta (\partial_x n)^2 \ .
\label{SM:eq-pert}
\end{equation}

For $\xi=0$, the analytical solution $n_0(x)$ that connects the
populated $n_+^*$ state for $x\rightarrow -\infty$
($n_0(x\rightarrow -\infty)=n_+^*$) with the bare-soil solution
when $x\rightarrow\infty$ ($n_0(x\rightarrow \infty)=0$) is
\begin{equation}
n_0 (x) = \frac{n^*_+}{2}\left(1-\tanh( \frac{n^*_+}{2}\sqrt{\frac{b}{2\epsilon}}x ) \right),\
\label{SM:front_solution}
\end{equation}
with velocity $v_0 = \sqrt{\frac{\epsilon b}{2}} (n^*_+ -
2n^*_-) $, where $n^*_\pm = (a\pm\sqrt{a^2-4\omega})/(2b)$.
Thus, writing $n = n_0 + \xi n_1 + \mathcal{O}( \xi^2)$ and $v
= v_0 + \xi v_1 + \mathcal{O}( \xi^2)$ and substituting in Eq.
(\ref{SM:eq-pert}), we obtain to first order in $\xi$:

\begin{align}
0 &= (-\omega +2an_0-3bn_0^2)n_1 \nonumber \\
  &+ \epsilon\partial_x^2 n_1 + v_0\partial_x n_1 + \delta (\partial_x n_0)^2 + v_1 \partial_x n_0.
\label{SM:first_order}
\end{align}

Defining the operator $\hat{\mathcal{L}} = (-\omega
+2an_0-3bn_0^2) +\epsilon\partial_x^2 + v_0\partial_x $ one can
write

\begin{equation}
\hat{\mathcal{L}}n_1 = - \delta (\partial_x n_0)^2 - v_1 \partial_x n_0.
\label{SM:first_order1}
\end{equation}

Let $\langle f\big|g\rangle =
\int_{-\infty}^{\infty} f(x)g(x)dx$ be the inner product,
$\hat{\mathcal{L}}^\dagger$ the adjoint of $\hat{\mathcal{L}}$,
and $w(x) = \exp(\frac{v_0}{\epsilon} x)\partial_x n_0$ the
neutral mode of $\hat{\mathcal{L}}^\dagger$
($\hat{\mathcal{L}}^\dagger w=0$). We can apply the solvability
condition $\langle w\big|\hat{\mathcal{L}} n_1\rangle = \langle
w\big|f(x)\rangle$, so that $\langle \hat{\mathcal{L}}^\dagger
w\big| n_1\rangle = \langle w\big|f(x)\rangle$ and $0 = \langle
w\big|f(x)\rangle$, to write
\begin{equation}
\langle w \big| - \delta (\partial_x n_0)^2 - v_1 \partial_x n_0 \rangle=0.
\label{SM:solvability}
\end{equation}

Thus, we can determine the contribution to the velocity due to
the effect of $\delta$:
\begin{equation}
v_1 = \frac{\langle w \big| - \delta (\partial_x n_0)^2 \rangle}{\langle w \big| \partial_x n_0 \rangle}.
\label{SM:solvability1}
\end{equation}

Solving the integrals in Eq. (\ref{SM:solvability1}), we obtain

\begin{equation}
v_1 =\frac{\delta\left(\frac{v_0^4}{\epsilon^2} - \frac{5 v_0^2 b n_+^{*2}}{2\epsilon} + b^2 n_+^{*4}\right)}{10\sqrt{2b\epsilon}\left(\frac{bn_+^{*2}}{2} -\frac{v_0^2}{\epsilon}\right)}.
\label{SM:velocity}
\end{equation}

The unperturbed velocity $v_0$ and the one with the first order
correction in $\delta$ ($v=v_0+v_1$, since $\xi=1$) are plotted
in Fig. (\ref{velocity}). The effect of $\delta$ is to increase
$v$ in the positive direction, i.e. to speed up the advance of
the populated state onto the unpopulated one.


%

\end{document}